\documentclass[twocolumn,showpacs,amsmath,pra,aps,amssymb,nofootinbib,superscriptaddress]{revtex4-1} 
\usepackage[T1]{fontenc}
\usepackage[latin9]{inputenc}
\usepackage{times}
\usepackage{color} 
\usepackage{xspace}
\usepackage{amssymb,amsmath}
\usepackage{amsbsy}
\usepackage[pdftex]{graphicx}
\usepackage{bm}
\usepackage{float}
\usepackage{adjustbox}

\usepackage[unicode,breaklinks]{hyperref}
\hypersetup{
    unicode=true,
    plainpages=false, 
    colorlinks=true,
    linkcolor=blue,
    citecolor=blue,
    filecolor=black,
    urlcolor=blue
}
\usepackage{url}
\usepackage{verbatim}

\newcommand{\UD}{U_{\mathrm{dd}}}
\newcommand{\UDt}{\tilde{U}_{\mathrm{dd}}}
\newcommand{\Cdd}{C_{\mathrm{dd}}}
\newcommand{\edd}{\epsilon_{\mathrm{dd}}}
\newcommand{\vtn}{V_\mathrm{tr}} 
\newcommand{\bk}{\mathbf{k}}
\newcommand{\br}{\mathbf{r}}
\newcommand{\bx}{\mathbf{x}}
\newcommand{\xic}{\xi^{\mathrm{crit}}}
\newcommand{\nt}{\tilde{n}}

\newcommand{\ldb}{\ensuremath{\lambda_{\mathrm{dB}}\xspace}}
\newcommand{\bose}[3][{}]{\ensuremath{\zeta^{#1}_{#2}\negthinspace\left(#3\right)}\xspace}
\newcommand{\bosee}[3][{}]{\ensuremath{\zeta^{#1}_{#2}\negthinspace\left(e^{#3}\right)}\xspace}
\newcommand{\pdiff}[2]{\dfrac{\partial #1}{\partial #2}}
\newcommand{\pdiffl}[2]{{\partial #1}/{\partial #2}}
\newcommand{\kt}{k_B T}
\newcommand{\aho}{\ensuremath{a_{\mathrm{ho}}}\xspace}
\newcommand{\veff}{V_{\mathrm{eff}}\xspace}
\begin{document}

\title{Stability of a trapped dipolar quantum gas}
\author{D.~Baillie} 
\affiliation{Dodd-Walls Centre for Photonic and Quantum Technologies, Department of Physics, University of Otago, Dunedin, New Zealand}
\author{R.~N.~Bisset} 
\affiliation{Center for Nonlinear Studies and Theoretical Division, Los Alamos National Laboratory, Los Alamos, New Mexico 87545, USA}
\author{P.~B.~Blakie}  
\affiliation{Dodd-Walls Centre for Photonic and Quantum Technologies, Department of Physics, University of Otago, Dunedin, New Zealand}

\begin{abstract}
We calculate the stability diagram for a trapped normal Fermi or Bose gas with dipole-dipole interactions (DDIs). Our study characterizes the roles of trap geometry and temperature on the stability using Hartree-Fock theory. We find that exchange appreciably reduces stability, and that, for bosons, the double instability feature in oblate trapping geometries predicted previously is still predicted by the Hartree-Fock theory. Our results are relevant to current experiments with polar molecules and will be useful in developing strategies to obtain a polar molecule Bose-Einstein condensate or degenerate Fermi gas.
\end{abstract}


\maketitle


\section{Introduction} 
There has been tremendous progress in producing ultra-cold  gases 
of atoms with strong magnetic dipoles \cite{Griesmaier2005a,Beaufils2008a,Lu2011a,Aikawa2012a,Lu2012a,Aikawa2014a} and heteronuclear  molecules with strong electric dipoles \cite{Ni2008a,Aikawa2010a,Takekoshi2014a}. The defining feature of these systems is that the  particles interact with a  DDI that is long-ranged and anisotropic. This anisotropy -- that  side-by-side dipoles repel, whereas those in a head to tail configuration attract -- significantly affects the properties of system. Notably, the attractive component of the interaction can cause the system to become mechanically unstable and collapse. Experiments using Bose-Einstein condensates have investigated the interplay between collapse and the geometry of the harmonic confinement potential \cite{Koch2008a} (also see \cite{Lu2011a,Muller2011a}) and the collapse dynamics of the system \cite{Metz2009a}. Theoretical understanding of condensate stability is well developed for the nearly pure condensate \cite{Ronen2007a,Lu2010a} (providing a good description of experimental results  \cite{Wilson2009a}) and the partially condensed system \cite{Bisset2012a}.
 
In contrast, the stability of normal Bose and Fermi gases are more poorly characterised. Such systems should be well described by Hartree-Fock theory, however these calculations are challenging in the presence of a confining potential, particularly near instability.   The Hartree (i.e.~direct interaction) term is reasonably convenient to implement in an efficient and accurate manner in calculations. This term (which is the only interaction term present in the Gross-Pitaevskii description of a dipolar condensate) tends to distort the position space density to elongate along the direction that the dipoles are polarised. The Fock term (i.e.~exchange interaction) is more difficult to implement, and for this reason is often neglected or treated approximately. This term tends to distort the momentum distribution in a manner that depends on the statistics of the particles \cite{Baillie2012a}, an effect that has recently been observed in a normal Fermi gas  \cite{Aikawa2014b}. A detailed understanding of stability is required by polar molecule experiments because of the large dipole moments that can be obtained with these molecules.  This understanding has to include both thermal and trapping effects, as to date these systems have not been cooled below the relevant degeneracy temperatures (i.e.~Fermi or condensation temperature), and high aspect ratio trapping potentials have been employed to suppress bimolecular chemical reaction rates \cite{Miranda2011a}.  From a theoretical perspective, it is important to establish whether transitions to novel manybody states will occur before mechanical instability (e.g.~see \cite{Baranov2004a,Lin2010a}). 

The first treatments of stability for the Fermi gas in a three-dimensional harmonic trap were based on Hartree  \cite{Goral2001b} and simplified variational Hartree-Fock \cite{Miyakawa2008a,Endo2010a} theories. Those results were shown to significantly overestimate the critical dipole strength at which the system became unstable by Zhang and Yi, who performed the first full Hartree Fock calculations \cite{Zhang2009a,Zhang2010a}. Instability in these calculations was identified by the numerical algorithm becoming unstable (e.g.~failure to converge and development of density spikes). 
Numerical instability is sensitive to grid choice and must be verified by careful convergence testing with refined grids. For the full Hartree-Fock calculations of a cylindrically symmetric trap, the system is represented by a four-dimensional distribution function, and the ability to refine the grid is constrained by computational resources. Thus a major outstanding issue is to produce reliable and accurate predictions for the Fermi stability.

Stability for the normal Bose gas was considered in Refs.~\cite{Bisset2011a,Bisset2012a} at the level of Hartree theory, and a variational Hartree-Fock theory \cite{Endo2011a}, which is inapplicable near the critical temperature. The Hartree calculations revealed that above, but close to the critical temperature, the Bose gas can have a novel double instability feature when confined in an oblate trap with the dipoles polarised along the tightly confined direction. Full calculations for the stability boundary of the normal Bose gas have not been performed at the level of Hartree Fock theory, and an important question is whether the  double instability feature survives when exchange interactions are included.

In this paper we develop a general theory of stability for trapped dipolar gases described by the Hartree-Fock approximation. We derive a result for the compressibility of the gas at trap centre that we use to identify the instability. 
We present results for the stability boundaries of Bose and Fermi gases as a function of temperature and trap geometry. Importantly, we show that these boundaries can be accurately computed, and show that previous Hartree-Fock results for a dipolar Fermi gas in an oblate trap significantly overestimate the stability region. Our Bose gas stability calculations are the first Hartree-Fock results for this system, and demonstrate that the double instability feature is robust to exchange interactions.
We develop useful analytic results to describe the behaviour of the stability boundaries.

\section{Theory}
\subsection{Hartree-Fock formalism\label{s:HFformalism}}
We consider a gas of dipolar particles polarised into a single internal state and confined in an arbitrary trapping potential
$\vtn(\bx)$.
These particles interact via a DDI of the form 
\begin{align}
\UD(\br)=\frac{\Cdd}{4\pi}\frac{1-3\cos^2\theta}{|\br|^3},\label{e:Udd}
\end{align}
where $C_{\rm{dd}}=\mu_0\mu^2_m$ for magnetic dipoles of strength $\mu_m$ and $d^2/\epsilon_0$ for electric dipoles of strength $d$, and $\theta$ is the angle between the dipole separation $\br$ and the polarization axis, which we take to be the $z$ direction. We do not consider the case with $\Cdd<0$, which could be obtained by rapidly rotating the dipoles \cite{Giovanazzi2002a}. For bosons, the particles also interact via a contact interaction of strength $g$,  where $g=4\pi a\hbar^2/m$, with $a$ the $s$-wave scattering length. 

Working in the grand canonical ensemble, and making a semi-classical approximation \cite{Goral2001a,Zhang2010a,Lima2010a,Baillie2010b}, the system is described by the Wigner function
\begin{align}
    W(\bx,\bk)=\frac{1}{\exp\left(\beta\left[\frac{\hbar^2k^2}{2m} + \veff(\bx,\bk) - \mu\right]\right)-\eta},\label{e:wigner}
\end{align}
where $\eta=1$ for bosons and $\eta=-1$ for fermions, $\mu$ is the chemical potential, and $\beta=1/k_BT$. This approximation furnishes a good description when the temperature is large compared to the trap level spacing  of the confining potential. For fermions this approximation can be applied at $T=0$\footnote{For fermions at $T=0$, $W(\bx,\bk) = \Theta\left[\mu-\frac{\hbar^2k^2}{2m} -\veff(\bx,\bk)\right]$ where $\Theta$ is the Heaviside step function.} as long as the Fermi energy is sufficiently large compared to the level spacing \cite{Zhang2009a}. 
For the case of bosons at $T<T_c$, where $T_c$ is the critical temperature \cite{Glaum2007a}, a condensate emerges, which is not described by the semi-classical Wigner function.  In the Hartree-Fock approximation 
\begin{align}
    \veff(\bx,\bk) &\equiv \vtn(\bx) + 2g n(\bx)+ \Phi_D(\bx)+\eta\Phi_E(\bx,\bk),
\end{align}
is the effective potential, dependent on both position and momentum, and recalling that  $g=0$ for spin-polarized fermions. The position density is given by     
\begin{align}
    n(\bx) &= \int \frac{d\bk}{(2\pi)^3} W(\bx,\bk),\label{e:nx}
\end{align}
and \cite{Baillie2010b,Baillie2012a}   
\begin{align}
 \Phi_D(\bx)&=\int d\bx'\,\UD(\bx-\bx')n(\bx'),\label{e:PhiD}\\
&= \int \frac{d\bk}{(2\pi)^3} e^{i\bk\cdot\bx}\UDt(\bk) \nt(\bk),  \label{e:PhiDmom} \\
\Phi_E(\bx,\bk)&=\int \frac{ d\bk'}{(2\pi)^3}\UDt(\bk-\bk')W(\bx,\bk'),\label{e:PhiE}
\end{align}
are the direct and exchange interaction terms, respectively and $\nt(\bk)$ is the Fourier transform of $n(\bx)$. In Eqs.~\eqref{e:PhiDmom} and \eqref{e:PhiE}, $\UDt$ is the Fourier transform of the dipole-dipole interaction, given by
\begin{align}
    \UDt(\bk)=\Cdd \left(\cos^2\theta_\bk-1/3\right),
\end{align}
where $\theta_\bk$ is the angle between $\bk$ and $k_z$.  
 
To find equilibrium solutions Eqs.~\eqref{e:wigner}-\eqref{e:PhiE} must be solved self-consistently subject to the additional constraint of atom number, i.e. $N=\int  d\bx\,n(\bx)$ fixed by adjusting the chemical potential. 
In practice solving the Hartree-Fock equations is time consuming and resource intensive because of the high dimensionality of the Wigner function and difficulties with accurately evaluating the exchange interactions. We do not discuss aspects of the numerical solution here, but note that
techniques for accurately solving these equations have been presented in \cite{Zhang2010a,Zhang2011a,Baillie2010b}.

\subsection{Hartree formalism\label{s:Hformalism}}
If the exchange term is neglected the effective potential loses its $\bk$ dependence, i.e.~$ \veff(\bx,\bk)\to \veff(\bx)$, where
\begin{equation}
V_{\mathrm{eff}}(\bx)\equiv V_{\mathrm{tr}}(\bx)+ 2g n(\bx)+\Phi_D(\bx).\label{Veff}
\end{equation}
In this case Eq.~(\ref{e:nx}) can be evaluated \cite{Baillie2010b,Bisset2011a}
\begin{equation}
n(\bx)=\frac{1}{\ldb^3}\zeta_{3/2}^\eta\left(e^{\beta[\mu-V_{\mathrm{eff}}(\bx)]}\right),\label{den}
\end{equation}
where 
\begin{equation}
\zeta_{\alpha}^{\eta}(z)=\sum_{j=1}^{\infty}\eta^{j-1}\frac{z^j}{j^\alpha},
\end{equation}
 is the polylogarithm function, and $\ldb=h/\sqrt{2\pi mk_BT}$. We have thus arrived at a Hartree theory for the system\footnote{More correctly Hartree for the DDI and Hartree-Fock for any contact interaction.}, which involves solving Eqs.~\eqref{e:PhiDmom}, (\ref{Veff}) and (\ref{den})  self-consistently. This theory is computationally much simpler because it does not require the evaluation of the full Wigner function.  

\subsection{Stability condition}
The mechanical stability of a system requires a finite positive compressibility $\kappa=n^{-2}(\partial n/\partial \mu)_T$. Thus we seek the divergence of the compressibility, and hence density fluctuations, to identify the critical point at which the system becomes unstable.  This approach has been applied to normal dipolar gases at the Hartree level using the RPA treatment of density fluctuations \cite{Bisset2011a}. 

In this subsection we derive a result for the compressibility within Hartree-Fock theory that we use to identify instability.
Neglecting the density prefactor,  the \textit{scaled} isothermal compressibility is
\begin{align}
    \pdiff{n(\bx)}{\mu} &= \int \frac{d\bk}{(2\pi)^3} \pdiff{W(\bx,\bk)}{\mu}.
        \end{align}
 The Wigner function derivative is obtained from Eq.~(\ref{e:wigner}) as     
\begin{align}
    \pdiff{W(\bx,\bk)}{\mu} &=  W_\mu(\bx,\bk)\label{e:Wderiv}\\
    &\times \left\{1-\left[2g-\frac{\Cdd}{3} +\eta\pdiff{\Phi_E(\bx,\bk)}{n(\bx)}\right] \pdiff{n(\bx)}{\mu}\right\},\notag
   \end{align} 
   where   
   \begin{equation}
        W_\mu(\bx,\bk)  \equiv \beta \bosee[\eta]{-1}{\beta\left[\mu -\frac{\hbar^2 k^2}{2m}- \veff(\bx,\bk)\right]},
     \end{equation}
is the derivative of $W$ with respect to $\mu$ at constant $\veff$.
For the term in braces in Eq.~(\ref{e:Wderiv}) we have used that $\Phi_E(\bx,\bk)$ is local in $\bx$ and in evaluating $\Phi_D(\bx)$, for stability considerations we take $\bk$ along the direction where $\UDt(\bk)$ is most attractive\footnote{Anisotropy of the DDI means that  the compressibility is directionally dependent  (c.f.~anisotropic speed of sound in a dipolar condensate \cite{Lima2012a,Bismut2012a} and superfluidity \cite{Ticknor2011a}). We take the most attractive direction for the purpose of analysing stability.}, i.e.~$\theta_\bk=\pi/2$, giving $\UDt(\bk) = -\Cdd/3$. Hence we obtain
 \begin{align}
    \pdiff{n(\bx)}{\mu} &= \frac{ n_\mu(\bx)} {1+[2g-\Cdd/3 -\Cdd \xi_\eta(\bx)]n_\mu(\bx) },\label{e:sccomp}
\end{align} 
where we have defined
 \begin{align}
    n_\mu(\bx) &\equiv \int \frac{d\bk}{(2\pi)^3} W_\mu(\bx,\bk),\\
     \xi_\eta(\bx) &\equiv -\eta\int \frac{d\bk}{(2\pi)^3} \frac{W_\mu(\bx,\bk)}{n_\mu(\bx)} \frac{\partial \Phi_E(\bx,\bk)}{\Cdd \partial n(\bx)},
\end{align}
where $\partial \Phi_E(\bx,\bk)/ \partial n(\bx)$ is derived in Appendix \ref{s:PhiEdn}. The \textit{bare scaled compressibility} $n_\mu(\bx)$ corresponds to the long wavelength limit of the bare density response function used in the RPA treatment [i.e.~$k\to0$ limit of $\chi_0(\mathbf{x},\mathbf{k})$, e.g.~see \cite{Mueller2000a,Bisset2011a}], while the \textit{exchange parameter} $\xi_\eta$ derived here is a key result of our work which is required to accurately account for the compressibility of the Hartree-Fock solution. 

For the gas to be stable, i.e.~have a finite positive compressibility, we require that [from (\ref{e:sccomp}) since $n_\mu(\bx)$ is always positive] %
\begin{align}
   1+[2g-\Cdd/3 - \Cdd \xi_\eta(\mathbf{x})]n_\mu(\mathbf{x}),\label{e:stabcond}
\end{align}
is positive definite at all positions $\mathbf{x}$. The critical point, where compressibility diverges, occurs when (\ref{e:stabcond}) equals zero at some position\footnote{For cases where the confining potential has a minimum, the gas has its peak density at this location, and this is where the system will first become unstable if the DDI strength is increased. }, i.e. 
\begin{align}
   1+[2g-\Cdd/3 - \Cdd \xi_\eta^{\mathrm{crit}}]n_\mu^{\mathrm{crit}}=0.\label{e:critHF}
\end{align}

Neglecting the exchange parameter $\xi_\eta$, this result reduces to the Hartree stability condition  \cite{Mueller2000a,Bisset2011a}
\begin{align}
   1+[2g-\Cdd/3]n_\mu^{\mathrm{crit}}=0,\label{e:critH}
\end{align}
where
\begin{align}
    n_\mu(\bx) = \frac{\beta}{\ldb^3} \bosee[\eta]{1/2}{\beta[\mu-\veff(\bx)]}.\label{e:nmuH}
\end{align}

\section{Stability based on density}
\subsection{Parameters and full numerical results}
Here we consider the stability of a dipolar gas  of specified peak density $n$.
It is convenient to introduce the dimensionless temperature $t\equiv\kt m /(\hbar^2n^{2/3})$ and interaction $c\equiv\Cdd n^{1/3} m/\hbar^2$ variables. These parameters completely characterise the Fermi gas, while for the Bose gas we also require  the  ratio of DDI to contact strength  $\edd=\Cdd/3g$.  That is, the dimensionless bare compressibility $\bar{n}_\mu(c,t,\edd)=n_\mu\hbar^2/n^{1/3} m$ and exchange parameter $\xi_\eta(c,t,\edd)$ are completely determined by these variables. 
In practice these functions can be obtained from a self-consistent solution for $W(\bx,\bk)$ in a  trap, from which $n_\mu(\bx)$ and $\xi_\eta(\bx)$ can be calculated, and then mapped to the dimensionless variables using the local density at each $\bx$.

\begin{figure}[t]
\begin{center}
    \includegraphics[width=3.4in]{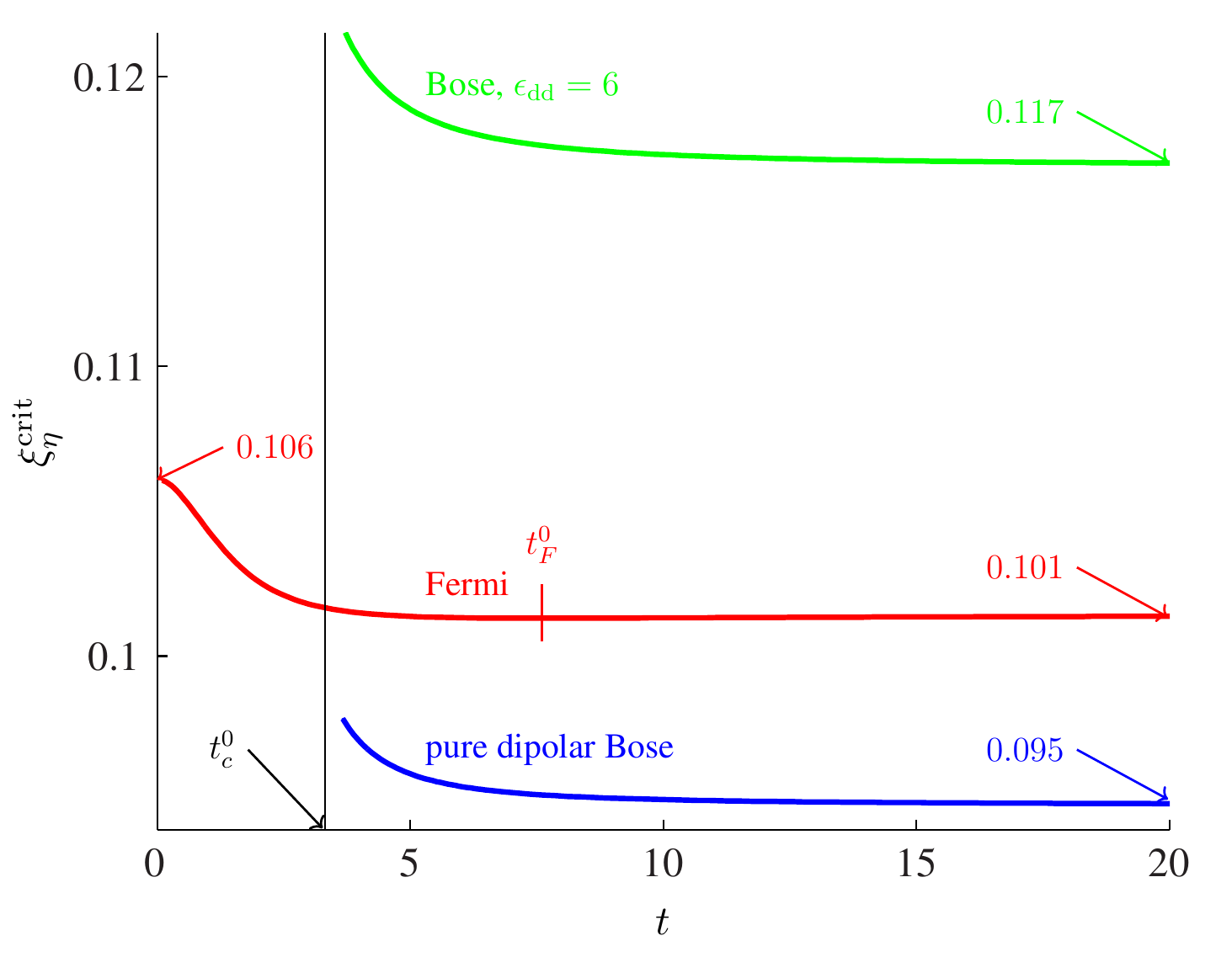} 
    \caption{(Color online) Critical exchange parameter $\xic_\eta$ for a Fermi gas (red), a purely dipolar Bose gas (blue) and a Bose gas with $\edd=6$ (green). 
    \label{f:xi}}
\end{center}
\end{figure}

From knowledge of functions $\bar{n}_\mu(c,t,\edd)$ and $\xi_\eta(c,t,\edd)$ we can determine the stability boundary as follows: at a given value of $t$ (and $\edd$ for bosons) the value of $c$ is increased until (\ref{e:critHF}) is satisfied. This identifies the critical interaction parameter $c^{\mathrm{crit}}$, exchange parameter (see Fig.~\ref{f:xi}), and bare compressibility.

\begin{figure}[t]
\begin{center}
    \includegraphics[width=3.4in]{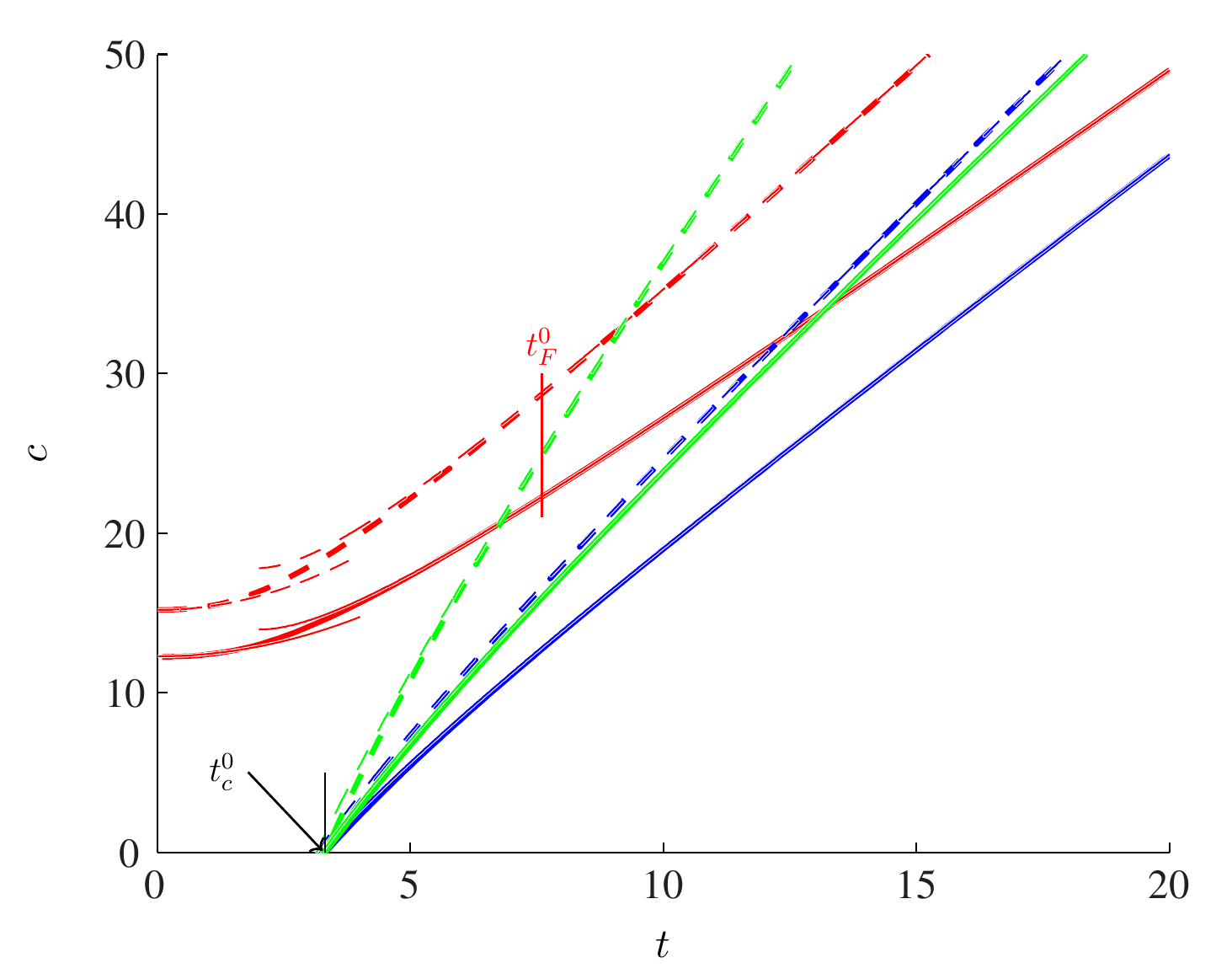} 
    \caption{(Color online) Stability boundaries based on the density of a Fermi gas (red curves), a purely dipolar Bose gas (blue) and a dipolar gas with $\edd=6$ (green) using Hartree-Fock theory (solid) and Hartree theory (dashed).  Thin lines are from Eqs.~\eqref{e:highTstab} for $t>2$ (almost obscured by the full results) and Eq.~\eqref{e:lowTstab} for $t<4$.
    \label{f:stabhfn}}
\end{center}
\end{figure}

The instability boundaries for Bose and Fermi gases using the Hartree-Fock theory are shown in Fig.~\ref{f:stabhfn}. For reference we have indicated the dimensionless non-interacting critical temperature for condensation $t_c^0 = 2\pi/\zeta(3/2)^{2/3} \approx3.31$ and Fermi temperature   $t_F^0=(9\pi^4/2)^{1/3}\approx 7.60$, for Bose and Fermi systems at  density $n$, respectively.
At $t_c^0$ the Bose gas saturates and for the purely dipolar case ($g=0$) is unstable for any non-zero value of the DDI (in the semiclassical approximation), whereas the Fermi gas, by virtue of its Fermi sea, is stable against a finite positive interaction down to zero temperature. The Fermi gas is always found to be stable against a larger dipole interaction parameter than a purely Bose gas at the same temperature. Our results for a  Bose gas with a $g>0$ ($\edd=6$) show that the $c^{\mathrm{crit}}$ boundary increases more rapidly with $t$, than for the purely dipolar case. For $g\ge\Cdd/3$ (i.e.~$\edd\le1$) the compressibility no longer diverges and the system is stable.

These observations are similar to those made in Ref.~\cite{Bisset2011a} using Hartree theory. For reference we have also included the Hartree boundary in  Fig.~\ref{f:stabhfn}, and observe that this always lies above the Hartree-Fock boundary. Thus we conclude that the effect of exchange interactions is to lower the stability boundary. 
Insight into this result can be found from the Hartree Local Fock (HLF) theory, an approximate Hartree-Fock theory introduced in Ref.~\cite{Baillie2012b}.  That treatment reduces the exchange interaction term to a local (i.e.~$\mathbf{x}$-dependent) form that is negative, but increases in magnitude with increasing $n(\bx)$. Thus the effect of exchange is qualitatively similar to an attractive contact interaction, acting to increase fluctuations and reduce stability.

\subsection{Analytic treatment}
The Hartree-Fock stability condition can be written [from Eq.~\eqref{e:critHF}] as
\begin{equation}
n_\mu^{\mathrm{crit}} =\frac{1}{\Cdd(1/3+ \xic_\eta-2/3\edd)}.\label{e:HFcrit}
\end{equation}

We develop analytic expressions for the local stability boundaries based on the HLF theory \cite{Baillie2012b} in which 
\begin{align}
    n &= (m^*/m)^{3/2}\ldb^{-3} \bose[\eta]{3/2}{z}, \label{e:nHLF}\\
    n_\mu &= (m^*/m)^{3/2}\beta\ldb^{-3}\bose[\eta]{1/2}{z}, \label{e:nmuHLF}
\end{align}
where $z$ is the local effective fugacity, $m^*$ is the local effective mass (approximately accounting for exchange effects), with $m/m^*=(1+2\delta/3)/(1+\delta)^{2/3}$, and the local momentum distortion, $\delta$, is discussed in Appendix~\ref{s:delta}. Using \eqref{e:HFcrit} we solve \eqref{e:nmuHLF} for the critical fugacity $z^{\mathrm{crit}}$, which we use in Eq.~\eqref{e:nHLF} to calculate the critical density to develop approximate analytic expressions for the critical interaction strength $c^{\mathrm{crit}}$.
\subsubsection{High $T$ limit}
In the limit $T\to\infty$ the exchange parameters tend  to the limits  indicated  in Fig.~\ref{f:xi}. 
 Using these constants we find
\begin{align}
    c^{\mathrm{crit}}  &\approx \frac{t - \eta  (\pi m/m^*)^{3/2} t^{-1/2} }{1/3 + \xic_\eta- 2/3\edd}, &   \label{e:highTstab}
\end{align}
for $t\gg1$, where, for  the values of $g$ we consider $m/m^*\approx 1.02$ to $1.03$, as can be found from $\delta$ in Appendix~\ref{s:delta}. This expression is plotted in Fig.~\ref{f:stabhfn}, and  found to provide a good approximation to the full Hartree-Fock results (thin solid lines almost obscured by the full results in the plot). Note that taking $m/m^*\approx 1$ does not noticeably change the results in Fig.~\ref{f:stabhfn} since it only applies to the $t^{-1/2}$ term. By setting the exchange effects to zero ($\xic_\eta=0$ and $m/m^*=1$) Eq.~(\ref{e:highTstab})  provides a good description of the full Hartree results (also shown in Fig.~\ref{f:stabhfn}).

\subsubsection{Low $T$ limit for fermions}
For fermions with $t \ll 1$, using the $T\to0$ limits for the exchange stability factor from Fig.~\ref{f:xi} and $m/m^*\to 1.07$ from Appendix~\ref{s:delta}
\begin{align}
    c^{\mathrm{crit}} &\approx \frac{(4\pi^4/3)^{1/3}m/m^*  + \frac{ 1}{9} (\pi/6)^{2/3} t^2 m^*/m}{1/3 + \xic_{-1}}.  \label{e:lowTstab}
\end{align}
This expression is plotted in Fig.~\ref{f:stabhfn}, and  found to provide a good approximation to the full numerical result.

\section{Stability of trapped gases }
In this section we specialize to particles confined within a cylindrically symmetric harmonic trap
\begin{align}
    \vtn(\bx)=\frac{m}{2}\left[\omega_\rho^2(x^2+y^2)+\omega_z^2z^2\right],
\end{align}
with aspect ratio $\lambda=\omega_z/\omega_\rho$. In contrast to the previous section we choose to focus on a system with fixed mean particle number $N$. The trap has a minimum at $\mathbf{x}=\mathbf0$ where the peak density of the gas occurs. Because the density is highest here, this location also determines the onset of instability according to when the density satisfies the critical condition presented in the previous section. Additional complexity for the fixed $N$ gas arises because the precise density at trap centre is determined by the interplay of the interactions and the trapping potential throughout the gas. For this case it is convenient to adopt the interaction parameter $D_t=\Cdd N^{1/6}/(4\pi\hbar\omega\aho^3)$, where $\omega$ is the geometric mean trap frequency and $\aho=\sqrt{\hbar/m{\omega}}$ (also used in \cite{Zhang2010a,Zhang2011a}).

\begin{figure}[t]
\begin{center}
    \includegraphics[width=3.4in]{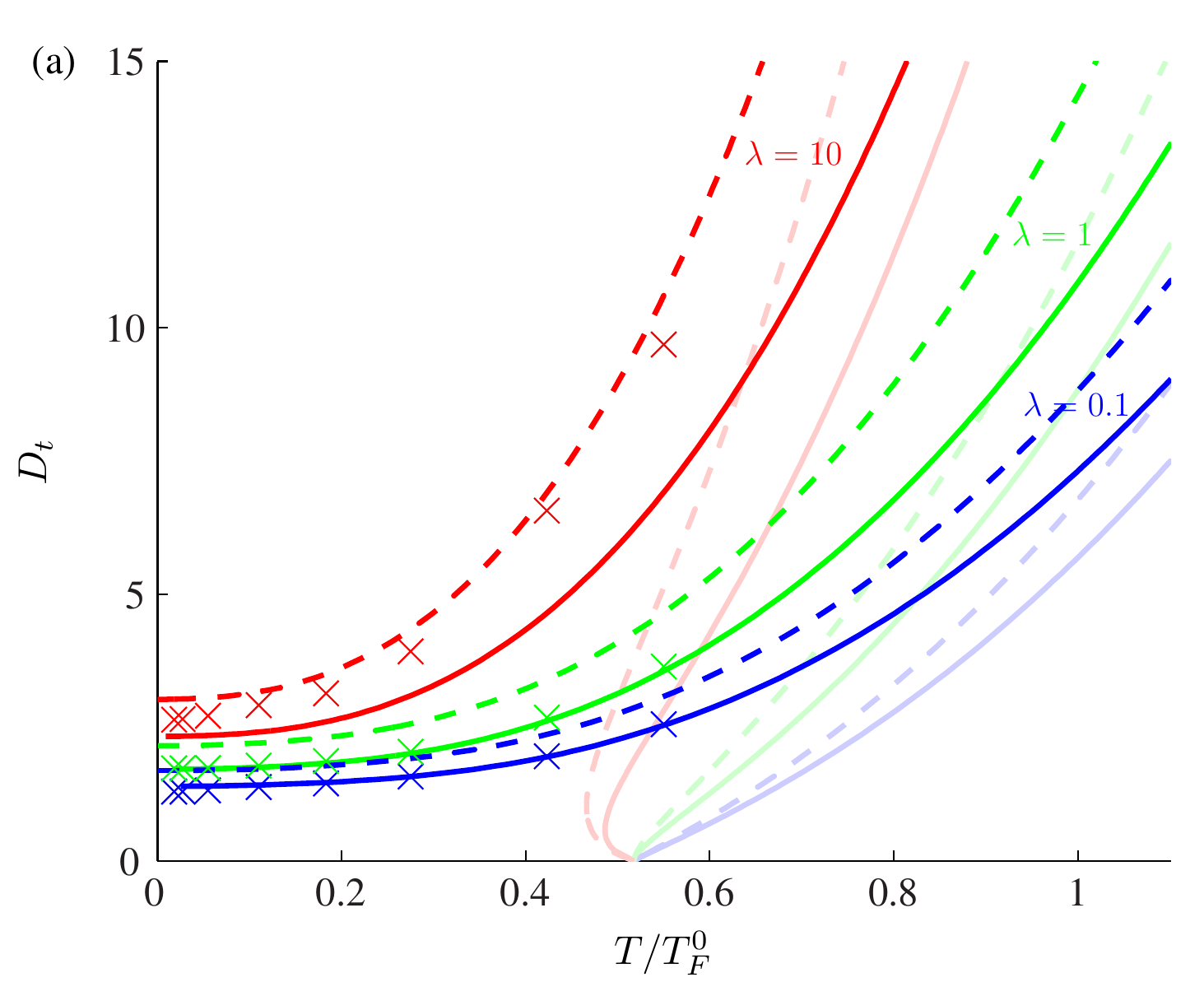} 
    \includegraphics[width=3.4in]{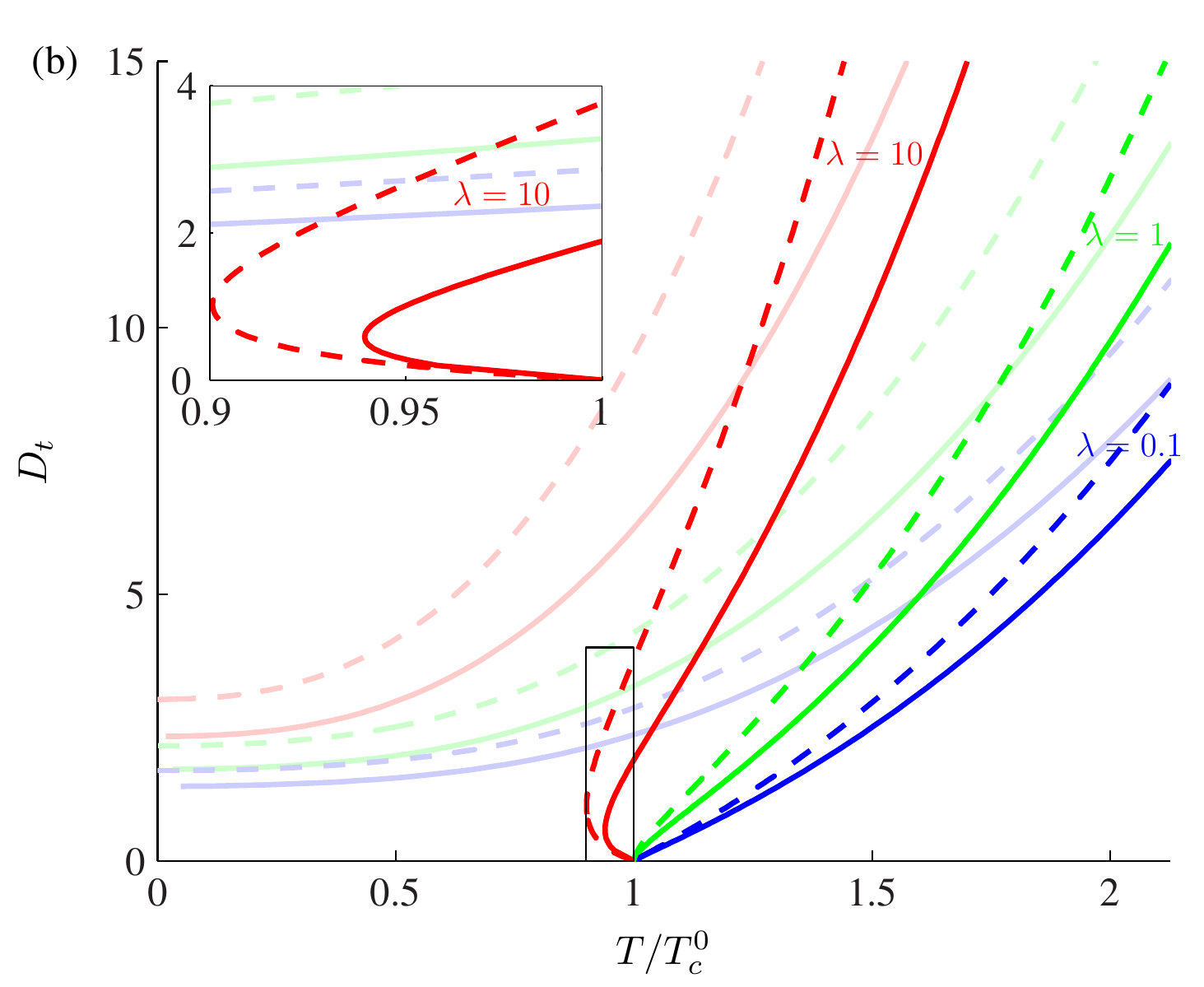} 
    \caption{(Color online) Stability boundaries of a (a) Fermi and (b) purely dipolar Bose gas with Hartree-Fock theory (solid) and Hartree theory (dashed) for $\lambda = 0.1$ (blue), $\lambda=1$ (green) and $\lambda=10$ (red). Corresponding results for a (a) Bose and (b) Fermi gas are shown as faint curves. We use $T_F^0=\hbar\omega(6N)^{1/3}/k_B$ and $T_c^0=\hbar\omega[N/\zeta(3)]^{1/3}/k_B$. Also shown are Hartree-Fock results from \cite{Zhang2010a} ($\times$). The inset to (b) enlarges the double instability region near $T_c^0$ marked by the box.
    \label{f:stabhfN}}
\end{center}
\end{figure}

In Fig.~\ref{f:stabhfN}(a) we show the results for the stability of a Fermi gas as a function of temperature at constant $N$ for prolate ($\lambda=0.1$), spherical ($\lambda=1$), and oblate ($\lambda=10$) trap geometries. These results demonstrate that stability increases with increasing aspect ratio. A simple interpretation is that for the oblate geometry the particles are mostly in a repulsive side-by-side configuration with respect to the DDI. This tends to expand the gas, reducing the central density and hence delaying the onset of instability. In contrast for the prolate trap the DDI is attractive and increases the central density. We have also plotted the Hartree-Fock stability predictions for these trap aspect ratios reported in \cite{Zhang2011a}. These results were determined by increasing the interaction parameter until the  procedure for calculating the self-consistent equilibrium state became numerically unstable, e.g.~forming density spikes. We find that these predictions significantly overestimate stability in the oblate trap compared to our results. In our own investigations into the use of numerical instability to locate the stability point, we find that the results are very sensitive to the numerical grids used for the calculations, particularly in oblate trap geometries.

In Fig.~\ref{f:stabhfN}(b) we show the results for the stability of a Bose gas as a function of temperature for the same set of trap aspect ratios. For these results we have restricted our attention to the purely dipolar interactions (i.e.~$g=0$). For the oblate case the stability boundary bends back on itself giving rise to a double instability region at temperatures $T<T_c^0$. However, the size of this region is smaller than the Hartree prediction.

\section{Conclusions}
In this paper we have used Hartree-Fock theory to predict the stability of harmonically trapped dipolar Bose and Fermi gases. We have developed an expression for the scaled compressibility of the Hartree-Fock solution, and use the divergence of this to identify the stability boundary. This allows us to predict stabilities boundaries from the Hartree-Fock solutions with high accuracy. Our results show that previous attempts using numerical instability significantly overestimated the stability boundaries in oblate trapping geometries.
 We have also applied our theory to provide the first Hartree-Fock stability predictions for the normal dipolar Bose gas. Importantly, we show that in an oblate trap the double instability features predicted in \cite{Bisset2011a} is maintained, although reduced in size. A number of analytic results for the stability boundaries are obtained and validated against the full numerical results.
\begin{figure}[t]
\begin{center}
    \includegraphics[width=3.4in]{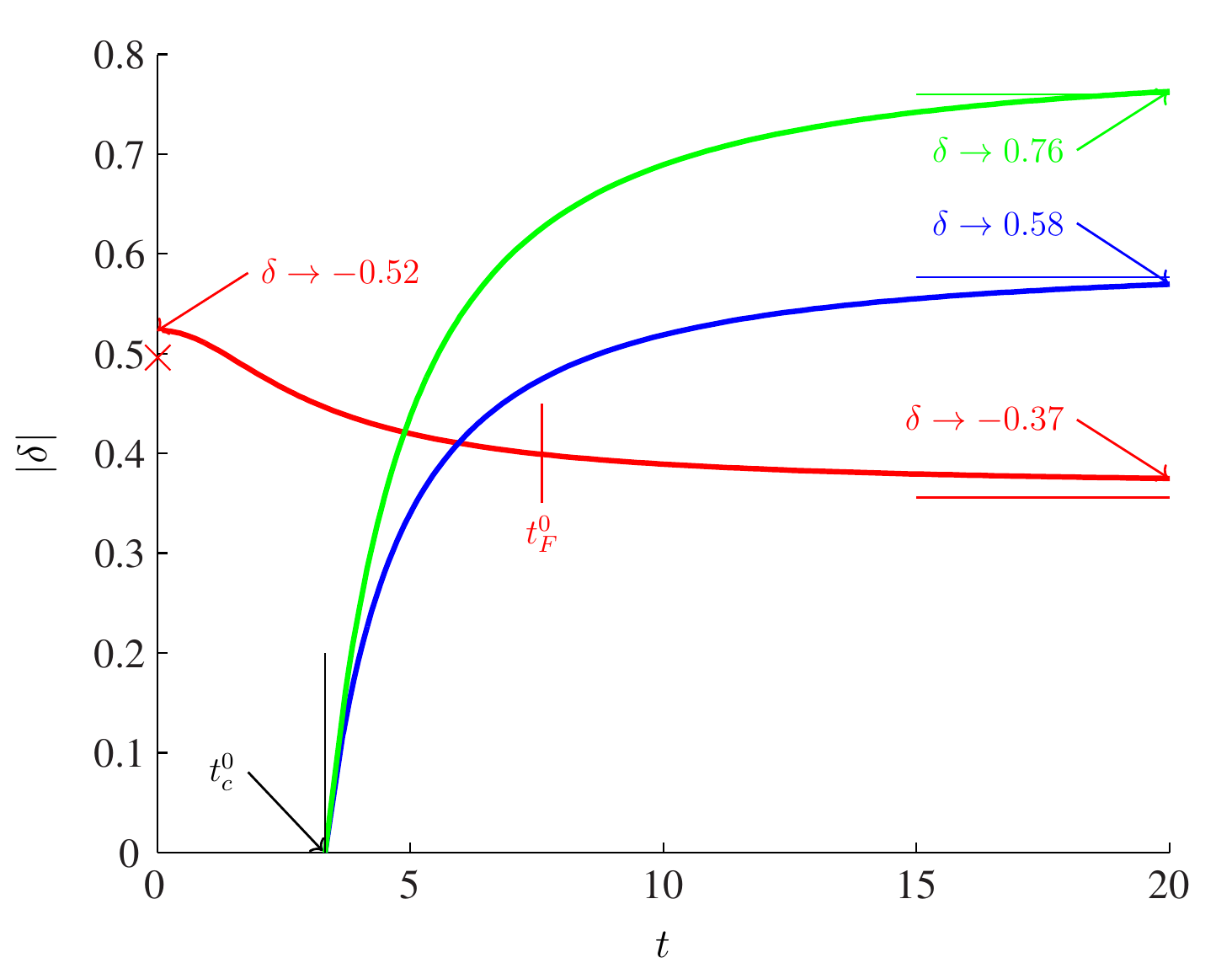} 
    \caption{(Color online) Absolute local momentum distortion at instability using Hartree-Fock theory of a  Fermi gas ($\delta<0$, red), a Bose gas ($\delta>0$), purely dipolar (blue) and with $\edd=6$ (green). Also shown are the HLF limits as $T\to\infty$ (thin lines) and $T\to 0$ ($\times$) showing qualitative agreement with the Hartree-Fock results.    
    \label{f:deltastab}}
\end{center}
\end{figure}

 Our results indicate that exchange interactions significantly shift the stability boundaries. At first sight this observation seems at odds with the rather small size of momentum distortion (arising from exchange) predicted for harmonically trapped Bose and Fermi gases, even near instability (see \cite{Baillie2012a} and \cite{Aikawa2014b}). However, it is worth emphasising that the momentum distortion arises from the entire system, including the many low density regions away from trap centre where exchange effects are small. In contrast stability is determined by the peak density part of the system, where locally exchange effects are strongest (Fig. 2(a) of \cite{Baillie2012b}). 

An interesting direction for future work would be to consider beyond semiclassical effects on stability, which would require diagonalizing for the low energy excitations (still treating the high energy excitations semiclassically see \cite{Bisset2012a,Ticknor2012a}). Also, the extension of the theory to the planar dipolar system, where one direction is tightly confined (e.g.~see \cite{Yamaguchi2010a,Zinner2011a}).

\appendix

\section*{Acknowledgments}
We gratefully acknowledge the contribution of NZ eScience Infrastructure (NeSI) high-performance computing facilities and support by the Marsden Fund of the Royal Society of New Zealand (contract number UOO1220).

\appendix
\begin{widetext}
\section{Derivation of $\pdiffl{\Phi_E }{n }$ \label{s:PhiEdn}}
Here we outline the basic steps used to derive our result for $\pdiffl{\Phi_E(\bx,\bk)}{n(\bx)}$.
First we note that
\begin{align}
 \pdiff{W(\bx,\bk)}{n(\bx)} &\equiv \pdiff{W(\bx,\bk)}{\mu} / \pdiff{n(\bx)}{\mu} 
 = \left[ 1-\Cdd \xi_\eta(\bx) n_\mu(\bx)\right]\frac{W_\mu(\bx,\bk)}{n_\mu(\bx)}-\eta W_\mu(\bx,\bk)\pdiff{\Phi_E(\bx,\bk)}{n(\bx)}. 
 \end{align}
 Using this result we obtain
\begin{align}
    \pdiff{\Phi_E(\bx,\bk)}{n(\bx)} &\equiv \pdiff{\Phi_E(\bx,\bk)}{\mu} / \pdiff{n(\bx)}{\mu}=  \int \frac{d\bk'}{(2\pi)^3} \UDt(\bk-\bk') \pdiff{W(\bx,\bk')}{n(\bx)},\\
&= \left[ 1-\Cdd \xi_\eta(\bx)n_\mu(\bx)\right]\int \frac{d\bk'}{(2\pi)^3} \UDt(\bk-\bk') \frac{W_\mu(\bx,\bk') }{n_\mu(\bx)}-\eta\int \frac{d\bk'}{(2\pi)^3} \UDt(\bk-\bk') W_\mu(\bx,\bk') \pdiff{\Phi_E(\bx,\bk')}{n(\bx)}.\label{e:dPhiEdn}
\end{align}
To solve the integral equation for $\pdiffl{\Phi_E(\bx,\bk)}{n(\bx)}$, we start with $\pdiffl{\Phi_E(\bx,\bk)}{n(\bx)}=0$ and iterate until self-consistent.
\end{widetext}

\section{Local momentum distortion at instability \label{s:delta}}
The local momentum distortion is defined as \cite{Baillie2012b} 
\begin{align}
 \delta(\bx) \equiv \frac{\gamma_{k_x}(\bx)}{\gamma_{k_z}(\bx)} -1, \hspace{2mm} \gamma_\nu(\bx) \equiv \int \frac{d\bk}{(2\pi)^3} \nu^2 W(\bx,\bk),
\end{align}
where $\gamma_{k_x}(\bx)$ and $\gamma_{k_z}(\bx)$ are the local momentum moments. The local momentum distortion at instability is shown in Fig.~\ref{f:deltastab}, where the numerical values for the  $T\to\infty$ limits for the Fermi and Bose systems are given, also the   $T\to 0$ limit for fermions. 

In the HLF theory at instability the local momentum distortion satisfies
\begin{align}
      \delta = -\eta \frac{c_t (1+2\delta/3)(1+\delta)J'(\delta)}{1/3+\xic_\eta-2/3\edd},
\end{align}
where $J$ is defined in \cite{Baillie2012b}, $c_t \to 3/2$ as $T\to \infty$ and $c_t\to 5/2$ for fermions as $T\to 0$ with results shown in Fig.~\ref{f:deltastab}.
%

%

\end{document}